\documentclass[showpacs, twocolumn]{revtex4-1}
\usepackage{graphicx}
\usepackage{amsmath}
\usepackage{appendix}
\usepackage{xcolor}
\usepackage{colortbl}

\begin{document}

\title{Modeling dark matter as self-bound quantum liquid droplets}

\author{Abdel\^{a}ali Boudjem\^{a}a }
\affiliation{Department of Physics, Faculty of Exact Sciences and Informatics.\\
Laboratory of Mechanics and Energy, Hassiba Benbouali University of Chlef, P.O. Box 78, 02000, Ouled-Fares, Chlef, Algeria.}
\email {a.boudjemaa@univ-chlef.dz}

\date{\today}

\begin{abstract}

The Bose-Einstein condensate dark matter model, where dark matter can be thought of as a non-relativistic, Newtonian gravitational condensate,
has recently attracted a great deal of interest.
In the present study, we explore the possibility that the dark matter could exist in the form of a self-bound quantum droplet 
formed by ultradilute quantum Bose mixtures under the action of Lee-Huang-Yang corrections at zero temperature.
To this end, we derive  an extended equation of state by using the nonrelativistic self-consistent  Hartree-Fock-Bogoliubov theory and the hydrodynamic approach.
The solutions of the obtained equations of state show that  the key parameters of the dark matter halos such as the density, mass, and radius are sensitive to the
interspecies interaction and to the quantum fluctuation strength.
The stability and the dynamical evolution of the droplet Galactic halos are analyzed by considering small perturbations of the quantum hydrodynamical equations.
In order to increase the reliability of our predictions we compare them with some observed data for the Galactic rotation curves.

\end{abstract}

\maketitle

\section{Introduction} \label{Intro}

Despite tremendous research on the nature of dark matter (DM), it remains a scientific mystery.
The detection of this mysterious material is challenging either locally or directly since  it interacts with ordinary baryonic matter and radiation only through gravity 
and thus, has extremely low density.

Many theoretical models have been proposed to describe  the formation and the behavior of the Galactic DM.
Among them one can quote the $\Lambda$-cold dark matter ($\Lambda$CDM) which suggests that dark matter moves nonrelativistically in the early Universe, 
and interacts only weakly (gravitationally) with ordinary matter and light (see for review \cite{Peeb1, Over,Weinb, Popo}). 
Although the CDM model works very well on large scales, it faces several issues such as  the "core-cusp problem" on smaller (Galactic) scales \cite{Moor}.
Such a catastrophe arises from the fact that DM is pressureless implying that gravitational collapse occurs at all scales and
the existence of CDM particles is yet inevident. 
Alternative ways to solve this dilemma is to consider self-interacting dark matter \cite{Sperg}, or the so-called warm DM \cite{Bode}, and hot DM \cite{Cald}.

Another intersting DM model known as Bose-Einstein Condensate  DM (BEC-DM), was first proposed in \cite{Sin} and further developed in \cite{Hu}.
In this model, DM halos are assumed to be composed of BECs and can be regarded as a giant bosonic system where the bosonic atoms 
are accumulated in a single macroscopic quantum state and described by the condensate wavefunction which  can be interpreted as a scalar field. 
This latter is crucial at small scales where it prohibits gravitational collapse, and solving small-scale problems \cite{PH}.
The dynamical and equilibrium properties of such a fuzzy DM have been deeply studied at both zero and finite temperatures using different approaches namely,
the Gross-Pitaevskii equation with hydrodynamic expansion, and Hartree-Fock-Bogoliubov (HFB) approximation (see \cite{Good, Peeb,Silv,Arb,Harko,Lee,Lee1,PH2,Harko1,Rind,Guz,Sharma} 
and references therein). 
In the framework of these theories, the DM can then be considered as a non-relativistic, Newtonian gravitational BEC,  
its pressure is related by a barotropic equation of state (EoS) \cite{Harko,Harko1}. 
One should stress also that the structure formation of binary BEC DM haloes has been attracted a wide interest in recent years (see e.g. \cite{Medv,Todo, Berman,Luu1, Pazo, Luu}).

Recently, ultracold atomic gases have offered an ideal platform  for creating a complex quantum many-body  beyond the mean-field approximation such as 
stable droplets in an ultradilute quantum regime \cite{Petrov}.
Experimentally, quantum droplets (QD) have been realized in single-component dipolar BECs \cite{Pfau,Pfau1,Chom} and in binary BECs \cite{Cab,Sem,Err}.
This intriguing new state of matter forms due to a balance of an attractive mean-field energy and repulsive beyond-mean-field effects originating 
from the  Lee-Huang-Yang (LHY) quantum fluctuations \cite{LHY}.
Self-bound QDs in ultracold Bose-Bose mixtures have been the focus of several recent theoretical studies owing to their fascinating properties 
(see e.g. \cite{Petrov1,Boudj,Sach, Cik1,Abbas1,Mehri}  and references therein).
An ultradilute QD close to zero temperature can be modeled using  the generalized Gross-Pitaevskii equation (GGPE) \cite{Petrov}. 
A finite-temperature version of the GGPE has been proposed in \cite{HHu, Ota, Boudj2} where the impacts of thermal fluctuations have been taken into account.

On the other hand, QDs have inspired to bridge the gap between atomic physics, condensed matter, quantum gravity, and cosmology.
Most recently, we have shown that QDs present a rare opportunity to probe the quantum nature of gravity \cite{Asma} and to 
test singularity resolution in black hole analogs \cite{Stanley}.
One common characteristic to QDs and CDM is the stabilization mechanism where both structures stabilize due to the delicate competition 
between attractive (collapse) and repulsive forces.  
Such a balance between an attractive self-interaction and a repulsive LHY term is closely analogous to the self-interaction transition
considered in the frame of the Coleman-Weinberg potential inspired by elementary particle physics \cite{CW, Moss}.
Futhermore, they are both ultradilute comoposites (weakly interacting systems).
The flat-top shape of QDs may lead to more robust DM against collapse instabilities than the usual Gaussian-like BEC.
Moreover, the absence of thermal reactions in DM due to electromagnetism makes it cold and coherent matter similarly to QD with its self-evaporation process 
(i.e. any excess of energy is expelled by losing particles) which also maintains it at zero temperature.
In addition to their longevity, another fascinating aspect of droplets is that they can keep their shape  during their evolution, 
render them important in cosmology, notably in the early Universe \cite{Must}. 
Other models for DM consisting of uniform droplets of BEC have been proposed in \cite{Moss,Harko2,Delgad, Panoto}.

The aim of the present paper is then to systematically explore the possibility that such an exquisite DM could exist in the form of binary BECs and of quantum droplet DM (QD2M).
We calculate in particular the density and mass profiles, the radius of DM, the rotation curve, and the velocity dispersion for both binary BEC DM and QDs.
To this end,  we derive the EoS using our Hartree-Fock-Bogoliubov (HFB) theory \cite{Boudj3,Boudj4,Boudj5}. 
This latter offers a highly controllable platform for investigating the properties of ultracold  Bose gases at both zero and finite temperatures\cite{Boudj3}
and has been validated extensively in many problems including collective modes \cite{Boudj6,Boudj7}, vortices \cite{Boudj8,Boudj9}, solitons \cite{Boudj10,Boudj11,Boudj12}, Bose mixtures 
\cite{Boudj, Boudj4,Boudj12,Boudj13}, self-bound quantum droplets \cite{Boudj,Boudj2,Abbas1}, and quantum gravity \cite{Asma,Boudj14}.  
The HFB theory includes self-consistently higher-order quantum and thermal fluctuations stemming from the noncondensed and anomalous densities into the EoS, 
generalizing the well-know GPE. 
Another feature of the HFB equations is that coupling between the condensate wavefunction (order parameter) and the noncondensed and anomalous densities enables one
to capture the behavior of ultracold Bose gases even beyond the dilute regime and at any temperature. 

In the case of a symmetric binary BEC DM,  the solution of the HFB EoS allows us to derive a generalized mathematical expression for the basic equations describing the
DM such as the density and mass profiles, the radius of DM, and the tangential velocity  obtained in terms of the interspecies interaction.
We show that such a self-gravitating Bose mixture is equivalent to a polytrope of index 1 with a correction term depending on the interspecies interactions.
A comparison between the velocity of the rotation curves predicted by our binary BEC DM model and the observational data of three galaxies 
makes our findings quantitative and credible.

A remarkable feature of the QD is that its wavefunction admits a hydrodynamical description implying that the dynamics of the system is governed the continuity and Euler equations,
facilitating the extension to the cosmological domain.
The exact solution of the resulting extended EoS is a difficult task thus, we critically resort to a numerical scheme which enables us to approximate solutions
for the density and mass profiles, and the tangential velocity.
Our results reveal that the additional pressure arising from the repulsive LHY quantum fluctuations may strongly stabilize 
the system against mean-field attraction and gravitational potential leading to a flat central density core droplet preventing  cuspy density profiles at galactic centers. 
We show that these  QD2M have universal mass profiles and corresponding universal rotation curves.

The stability of the observed self-bound cosmological droplets under small perturbations in the homogeneous case (i.e. in the flat-top region) is also studied, 
and the gravitational Bogoliubov spectrum is obtained.
Furthermore, we accurately examine the time evolution of QD halos  using the super-Gaussian variational ansatz, offering insight into equilibrium state, the frequency of the oscillation modes, 
and the dynamics of the system radius. 
Finally, we apply the obtained formulas on a galaxy made up of a QD2M halo in order to test the reliability of our findings with some observed galactic rotation curves. 
We find an excellent agreement between our theoretical approach and the observed data.

The rest of the paper is organized as follows.
In Sec.~\ref{Mod}, we outline the HFB formalism that describes the properties of binary Bose mixtures and derive the extended EoS in the presence of the LHY quantum corrections.
In Sec.~\ref{MixDM} a Newtonian dark matter binary BEC is proposed and analyzed.
Section \ref{SBDM} addresses the formation, the stability, and the dynamics of QD2M. 
The key parameters of DM halos such as the mass, the radius, the effective potential, and frequency of the breathing modes are computed,
within the realm of the variational approach.
We conclude our results in Sec.~\ref{concl}.

\section {Hartree-Fock-Bogoliubov theory: Equation of state} \label{Mod}

We consider a weakly interacting two-component BEC  with an atomic mass $m_j$ in a box of volume $\mathcal V$.
The many-body Hamiltonian describing  such a system can be written as:
\begin{align} \label{eq4}
\hat H &= \sum_{j=1}^2\int d{\bf r} \, \hat\psi_j^\dagger ({\bf r}) \bigg[ h^{\text{sp}}+\frac{g_j}{2} \hat \psi_j^\dagger ({\bf r})\hat \psi_j ({\bf r})\bigg]\hat\psi_j ({\bf r}) \nonumber\\
&+g_{12}\int d{\bf r} \, \hat\psi_2^\dagger ({\bf r}) \hat \psi_2({\bf r})\hat \psi_1^\dagger ({\bf r}) \hat\psi_1({\bf r}),
\end{align}
where  $\hat\psi_j^\dagger$ and  $\hat\psi_j$ are the boson destruction and creation field operators, respectively, satisfying the usual canonical commutation rules 
$[\hat\psi_j({\bf r}), \hat\psi_j^\dagger (\bf r')]=\delta ({\bf r}-{\bf r'})$,  $h^{\text{sp}}=- (\displaystyle\hbar^2/\displaystyle 2m_j) \nabla^2+V(\mathbf r)$
is the single particle Hamiltonian, and $V(\mathbf r)$ is an external (which will be defined below).
The coefficients $g_j=(4\pi \hbar^2/m_j) a_j$ and $g_{12}=g_{21}= 2\pi \hbar^2 (m_1^{-1}+m_2^{-1}) a_{12}$ with 
$a_j$ and $a_{12}$ being the intraspecies and the interspecies scattering lengths, respectively. \\
The corresponding energy functional ${\cal E}=\langle H\rangle$ is given by
\begin{align}  \label{egy}
{\cal E}& = \sum_{j=1}^2 \bigg \{ \int d{\bf r} \, \left[\Phi_j^*  h^{\text{sp}} \Phi_j 
+ \hat{\bar \psi}_j^\dagger \bigg( - \frac{\displaystyle\hbar^2}{\displaystyle 2m_j} \nabla^2-\mu_j\bigg) \hat{\bar \psi}_j +\frac{g_j}{2}  n_j^2 \right] \bigg\} \nonumber\\
&+ g_{12} \int d{\bf r}  n_1n_2 + {\cal E}_{\text {LHY}}, 
\end{align}
where  $\hat{\bar \psi}_j({\bf r})=\hat\psi_j({\bf r})- \Phi_j({\bf r})$ is the noncondensed part of the field operator 
with $\Phi_j({\bf r})=\langle\hat\psi_j({\bf r})\rangle$. \\
The last term in Eq.~(\ref{egy}) accounts for the LHY correction to the energy which is included self-consistently. It is defined as: 
\begin{align} \label{LHY}
{\cal E}_{\text{LHY}} &= \frac{1}{2}\sum_{j=1}^2 g_j \int d{\bf r} \big( 2\tilde n_j n_j-\tilde n_j^2 +|\tilde m_j|^2 \\
&+ \tilde m_j^*\Phi_j^2+ \tilde m_j {\Phi_j^*}^2 \big), \nonumber
\end{align}
where $\tilde n_j=\langle\hat{\bar {\psi}}_j ^\dagger\hat{\bar {\psi}}_j\rangle$ is the noncondensed density, 
$\tilde m_j= \langle\hat {\bar {\psi}}_j\hat{\bar {\psi}}_j\rangle$ is the anomalous density,  $n_{cj}=|\Phi_j|^2$ is the condensed density, and
$n_j=n_{cj}+\tilde n_j$ is the total density.

The equation of motion for the condensate can be straightforwardly derived through $i\hbar \partial \Phi_j/\partial t =\partial{\cal E}/\partial \Phi_j^*$ (see \cite{Boudj3} and reference therein):
\begin{align} \label{GGPE}
i\hbar \dot{\Phi}_j & = \left[- \frac{\displaystyle\hbar^2}{\displaystyle 2m_j} \nabla^2+V(\mathbf r) +g_j (n_{cj}+2\tilde n_j) + g_{12}  n_{3-j} \right]\Phi_j \nonumber\\
&+ g_j\tilde m_j \Phi_j^{*},
\end{align}
For $g_{12} =0$ and in the absence of the gravitational field, we recover the standard time-dependent HFB equations \cite {Boudj3, Boudj6,Boudj7,Boudj8,Boudj9, Asma, Boudj14} 
describing a single-component BEC.

The  Bogoliubov excitation dispersion in the homogeneous mixture  can be computed by linearizing Eq.~(\ref{GGPE}) utilizing the generalized random-phase approximation \cite{Boudj6}: 
$\Phi_j = \sqrt{n_{cj}}+\delta \Phi_j $, $\tilde n_j=\tilde n_j+\delta \tilde n_j$,  and $\tilde m_j=\tilde m_j+\delta \tilde m_j$, 
where $\delta \Phi_j (\mathbf r,t)= u_{jk}  e^{i  {\bf k \cdot r}-i\varepsilon_k t/\hbar}+v_{jk} e^{i { \bf k \cdot r}+i\varepsilon_k t/\hbar} \ll \sqrt{n_{cj}}$, 
$\delta \tilde n_j \ll \tilde n_j$, and $\delta \tilde m_j \ll \tilde m_j$ \cite{Boudj6}. 
Keeping only second-order terms, we find that  the Bogoliubov excitation spectrum comprises an upper branch, $\varepsilon_{k+}$, and a lower branch: $\varepsilon_{k-}$:
\begin{equation} \label {Bog}
\varepsilon_{k\pm}= \sqrt{E_k^2+2E_k \mu_{\pm}},
\end{equation}
where $E_k=\hbar^2k^2/2m$, $\mu_{\pm}=  \frac{\bar g_1 n_{c1}} {2} f_{\pm} (\Delta, \alpha)$, $f_{\pm} (\Delta, \alpha)= 1 + \alpha \pm \sqrt{ (1-\alpha)^2 
+4 \Delta ^{-1}\alpha }$, $\alpha= \bar g_2 n_{c2}/\bar g_1 n_{c1}$, and $\Delta=\bar g_1\bar g_2/g_{12}^2$ is the miscibility parameter.
Here the density-dependent coupling constant  $\bar g_1=g_1(1+\tilde m/n_c)$ is introduced in order to circumvent the ultraviolet divergence in 
anomalous density and in the ground-state energy  (see below) \cite{Boudj, Boudj4,Griffin}. It leads also to a gapless Bogoliubov spectrum \cite{Griffin}. \\
In the limit $k \rightarrow 0$, the total dispersion is phonon-like
\begin{equation} \label{sound}
\varepsilon_{k \pm}= \hbar c_{\pm} k,
\end{equation} 
where the sound velocity associated with  the upper branch, $c_{+}$, and lower branch, $c_{-}$, reads as:
 \begin{equation} \label{sound1}
c_{\pm} ^2=\frac{1}{2} \left[ c_1^2+c_2^2 \pm \sqrt{ \left( c_1^2-c_2^2\right) ^2 + 4 \Delta^{-1} c_1^2 c_2^2} \right], 
\end{equation}
with $c_j = \sqrt{\bar g_j n_{cj}/m}$ being the sound velocity of a single BEC.
For $g_{12}^2> \bar g_1\, \bar g_2$, the spectrum (\ref{Bog}) becomes unstable and thus, the two condensates spatially separate.

At zero temperature, the noncondensed and anomalous densities can be given as \cite{Boudj,Boudj4}:
\begin{equation}\label {nor}
\tilde n_{\pm}=\frac{1}{2}\int  \frac{d^3 k} {(2\pi)^3}  \left[\frac{E_k+ \mu_{\pm}} {\varepsilon_{k \pm}}-1\right],
\end{equation}
and
\begin{equation}\label {anom}
\tilde m_{\pm}=-\frac{1}{2}\int  \frac{d^3 k} {(2\pi)^3}  \frac{ \mu_{\pm} } {\varepsilon_{k \pm}},
\end{equation}
This equation clearly shows that the anomalous density is crucial in the stability and the dynamics of Bose gases.

In the realm of the HFB theory, the quantum pressure of the weakly-interacting Bose mixture  is given by \cite{Boudj4}:
\begin{equation}\label {ThPr}
P=- \left(\frac{\partial E}{ \partial {\cal V}}\right)_T,
\end{equation}
where
\begin{align}\label{gerny}
E=E_0+\frac{1}{2} \sum_{\pm}  \int \frac{d^3 k} {(2\pi)^3}  \left(\varepsilon_{k \pm} - 2E_k-\mu_{\pm} \right),
\end{align}
is the ground-state energy with
$E_0= g ( n_{c}^2+ 4 n_{c} \tilde n +2\tilde n^2 +\tilde m^2 + 2 n_{c} \tilde m) +g_{12}  n^2$  being the mean-field energy.
The second term in Eq.~(\ref{gerny}) represents the LHY quantum corrections. It can be computed using the dimensional regularization 
to overcome the ultraviolet divergence arising from a naive treatment of the contact interaction.

A straightforward calculation yields for the total pressure:
\begin{align}\label{ThPr1}
P=& \frac{1}{2} \sum_{j=1}^2 g_j n_j^2 + g_{12}  n_1 n_2 +\frac{16  g_1 \sqrt{a_1^3/\pi}}{15\sqrt{2}} n_1^{5/2} \\
&\times \bigg(1+\frac{\tilde m_1-\tilde n_1}{n_1}\bigg)^{5/2}  \sum_{\pm} f_{\pm}^{5/2}(\Delta,\alpha).  \nonumber 
\end{align}
Equation (\ref{ThPr1}) generalizes naturally many EoS existing  in the literature.
For $g_{12}=0$, the EoS (\ref{ThPr1}) reduces to that of a single BEC DM \cite{Sharma}. 
The LHY correction appears in the last term of Eq.~(\ref{ThPr1}) and may crucially influence the structural properties (mass and radius) of binary BEC DM (see below). 

At zero temperature, Eqs.~(\ref{nor}) and (\ref{anom}) provide useful expressions for the total noncondensed, $\tilde n=\sum_{\pm} \tilde n_{\pm}$, and anomalous, 
$\tilde m =\sum_{\pm} \tilde m_{\pm}$, densities \cite{Boudj,Boudj2,Boudj4}
\begin{align}\label {NCD}
\tilde n = \frac{2\sqrt{2} }{3} \,n_{1} \sqrt{ \frac{n_{1} a_1^3}{\pi}} \bigg(1+\frac{\tilde m_1-\tilde n_1}{n_1}\bigg)^{3/2} \sum_{\pm} f_{\pm}^{3/2} (\Delta,\alpha),
\end{align}
and 
\begin{align}\label {MCD}
\tilde m =2\sqrt{2} \,n_{1}  \sqrt{ \frac{n_{1} a_1^3}{\pi}} \bigg(1+\frac{\tilde m_1-\tilde n_1}{n_1}\bigg)^{3/2} \sum_{\pm} f_{\pm}^{3/2}(\Delta,\alpha).
\end{align}
For $g_{12}=0$, Eqs.(\ref{NCD}) and (\ref{MCD}) reduce to those obtained for a single component BEC.
The expression of the anomalous density (\ref{MCD}) has been indeed obtained using the dimensional regularization \cite{Boudj,Boudj2,Boudj3} to overcome the ultraviolet divergence.
The solution of the self-consistent Eqs.~(\ref{ThPr1})-(\ref{MCD}) requires the use of an iterative method.

In next the sections, we apply our HFB findings to construct binary Bose mixture and quantum droplet dark matter models.

\section{Binary BEC dark matter} \label{MixDM}

Here we explore the possibility that DM could exist in the form of a binary BEC.
Motivated by recent experiments \cite{Cab}, we consider for simplicity symmetric Bose condensates with equal intraspecies coupling constants $g_1=g_2=g$, 
and equal densities in each component  $n_1=n_2=n/2$, $\tilde m_1=\tilde m_2=\tilde m/2$, and $\tilde n_1=\tilde n_2=\tilde n/2$.
We also neglect higher-order effects i.e. $\tilde n/n \approx \tilde m/n \ll1$, since they are less important in dilute ultracold boson systems.

Therefore, the above EoS (\ref{ThPr1}) reduces to 
\begin{align}\label{ThPr2}
P (n)=&  \left(\frac{\delta g}{g}\right)_+ g n^2 +\frac{16  g \sqrt{a^3/\pi}}{15\sqrt{2}} n^{5/2} \sum_{\pm} 4 \sqrt{2} \left(\frac{\delta g}{g}\right)_{\pm}^{5/2},
\end{align}
where $\left(\delta g/g\right)_{\pm} =1\pm g_{12}/g$.
This EoS allows one to find the density of profiles of the binray BEC DM.

Let us assume a smoothly varying density, it is then legitimate to treat  the local density as a uniform bosons gas; $n(r=0)=n_0$ and $dn(r=0)/dr=0$.
Working on spherically symmetric representation \cite{Matos} and neglecting the baryonic effects, the static profiles can be given by $\nabla P(r)= -mn(r) \nabla V$,
where $V=V_g$ is the gravitational potential which is determined self-consistently using the Poisson equation $\nabla^2V_g=4\pi Gm n(r)$. 
Under these assumptions we obtain the following hydrostatic equation:
\begin{equation} \label{DMEoS}
\frac{1}{r^2} \frac{d}{dr} \left[\frac{r^2}{n(r)}\frac{d P(n)}{dr}\right]= -4\pi G m^2n(r).
\end{equation}
In the absence of quantum corrections one has $P = \left(\delta g/g\right)_+ g n^2 $,  thus Eq.~(\ref{DMEoS}) reduces to
a polytropic equation with index $n=1$:
\begin{equation} \label{DMEoS1}
\frac{d}{dx} \left(x^2\frac{d \bar n}{dx}\right)+ x^2 \bar n=0,
\end{equation}
where $\bar n=n/n_0$, $n_0$ is the central density, and  $x= \sqrt{1/\big[2\left(\delta g/g\right)_+\big]}q r$, where $q=\sqrt{m^3 G/ \hbar^2 a}$.
Applying the boundary conditions: $\bar n_0=1$ and $\bar n'_0=0$, gives the exact solution for DM density \cite{Harko}:
\begin{equation} \label{DMDp}
n(r)\equiv n_{\text{DM}}(r)= n_0\frac{\sin\left(\sqrt{1/\big[2\left(\delta g/g\right)_+\big]}qr\right)}{\sqrt{1/\big[2\left(\delta g/g\right)_+\big]}qr},
\end{equation}
Having obtained the density, we can now calculate the mass profile binary BEC Galactic halo which is 
defined as: $m_{\text{DM}}(r)= 4\pi \int_0^{r} r'^2  n_{\text{DM}}(r') dr'$. After a simple integration, the mass can be expressed as:
\begin{align} \label{DMMs}
 m_{\text{DM}}(r)&= \frac{8\pi \left(\delta g/g\right)_+n_0 r}{q^2} \bigg[\frac{\sin\left(\sqrt{1/\big[2\left(\delta g/g\right)_+\big]}qr\right)}{\sqrt{1/\big[2\left(\delta g/g\right)_+\big]}qr} \nonumber\\
& -\cos\left(\sqrt{1/\big[2\left(\delta g/g\right)_+\big]}qr\right) \bigg].
\end{align}
Near the boundary of the halo, the binary condensates vanish, $n(R_{\text{DM}}) = 0$ and $\sqrt{1/\big[2\left(\delta g/g\right)_+}qr= \pi$. 
This yields for the radius of the condensates DM:
\begin{equation} \label{DMRad}
R_{\text{DM}}=  \pi \sqrt{2\left(\delta g/g\right)_+ \hbar^2 a / m^3 G}.
\end{equation}
The total mass of the binary BECs is defined as  $M_{\text{DM}}=m_{\text{DM}} (R_{\text{DM}})$. This gives:
\begin{equation} \label{DMM}
M_{\text{DM}}= \frac{\pi}{4} n_0 R_{\text{DM}}^3.
\end{equation}
We argue that the density, the radius, and the mass of the  binary BEC DM strongly depend on the interspecies interaction.

The tangential velocity of test particles in stable circular orbits, $ \dot{r}=0$, must satisfy the equation \cite{Matos}: 
\begin{equation} \label{Vel}
v_{\text{tg}}^2(r)= G m_{\text{DM}}(r)/r.
\end{equation}
Upon substituting Eq.~(\ref{DMMs}) into (\ref{Vel}), one gets for the circular velocity:
\begin{align} \label{Vel2}
v_{\text{tg}}^2(r)&= 2 \left(\delta g/g\right)_+ v_R^2 \Bigg[\frac{\sin\left(\sqrt{1/\big[2\left(\delta g/g\right)_+\big]} \pi r/R_{\text{DM}}\right)}
{\sqrt{1/\big[2\left(\delta g/g\right)_+\big]} \pi r/R_{\text{DM}} } \nonumber\\
& -\cos\left(\sqrt{1/\big[2\left(\delta g/g\right)_+\big]} \pi r/R_{\text{DM}}\right) \Bigg],
\end{align}
where $v_R = \sqrt{G M_{\text{DM}}/R_{\text{DM}}}$ is the tangential velocity at the boundary $r=R_{\text{DM}}$.
For $r > R_{\text{DM}}$, we reproduce the standard Keplerian law.

Evidently, Eqs.~(\ref{DMDp})-(\ref{Vel2}) provide a full description of the structural properties of the binary BEC dark matter in terms 
of the interspecies interaction scattering length, the central density and the mass of the condensed atom. 
For $g_{12}=0$,  they reduce to their standard  form of a single BEC DM \cite{Harko,Harko1, PH2}.

To be concrete, we consider a galactic DM halo of mass $M_{\text{DM}}= 10^{11}M_{\odot} = 2\times 10^{44}$ g, extending with a radius $R_{\text{DM}}=10$ kpc $=3.08\times 10^{22}$cm. 
In the terrestrial experiments performed with two ${}^{39}$K BECs, the scattering length of each component is $a \approx 4.71\times 10^{-7}$ cm  $\approx 4.71\times 10^{6}$ fm \cite{Cab,Sem}.
For realistic galactic DM halos, the central density of the DM is in the range of $n_0 \approx 10^{-24}-10^{-26}$ g/cm${^3}$.
Therefore, the Newtonian approximation can be safely adopted for the analysis of the astrophysical properties of the binary BEC DM.

\begin{figure*}
	\includegraphics[scale=0.55]{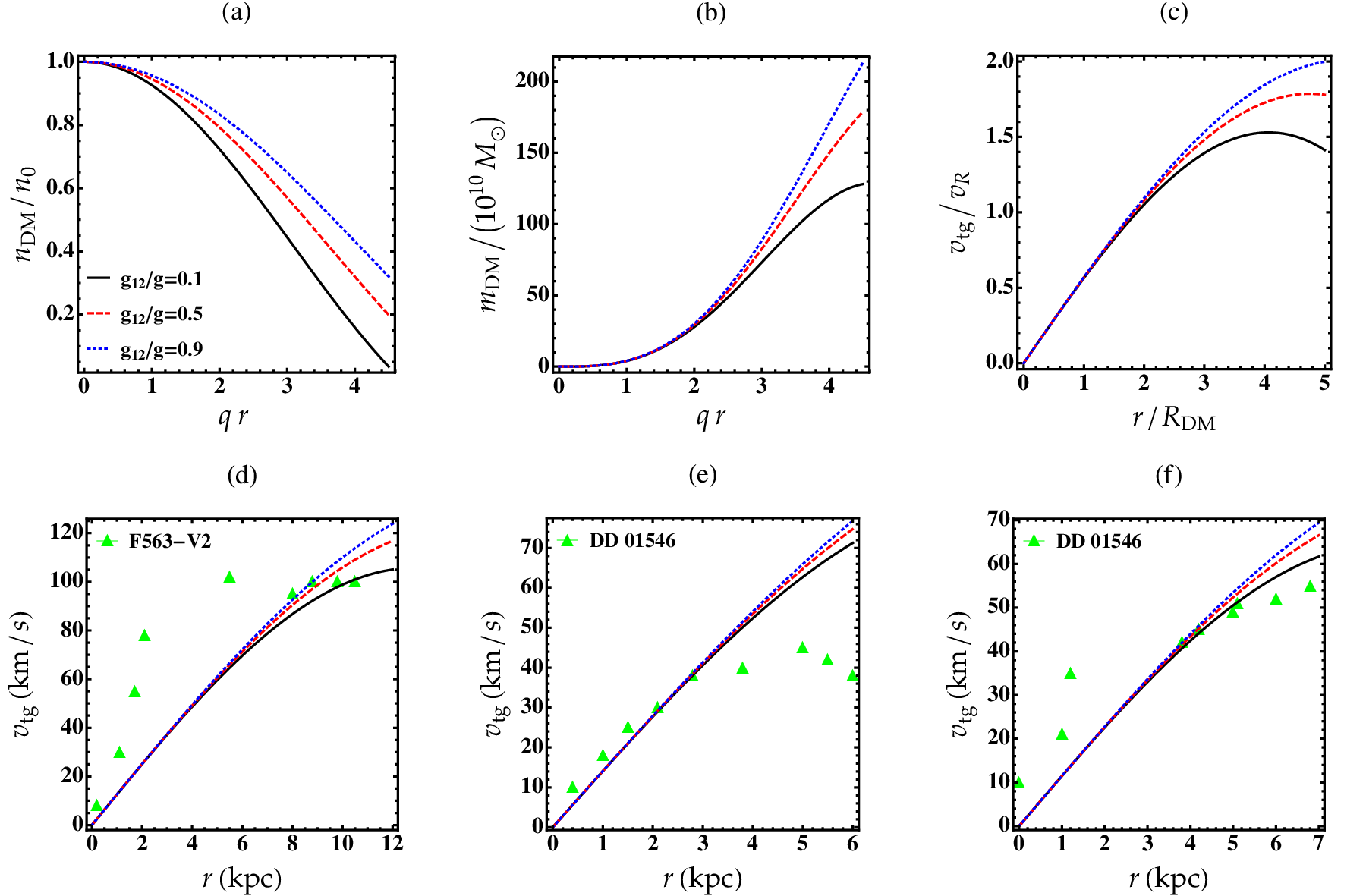}
		\caption{(a) Density profile in units $n_0$ and (b) mass distribution of a binary BEC DM for different values of the interspecies interaction.
(c) Normalized tangential velocity, $v_{\text{tg}}/v_R$ as a function of the radial distance $r/R_{\text{DM}}$, from Eq.~(\ref{Vel2}) of a galaxy consisting 
of a binary BEC DM halo for different values of the interspecies interaction. 
(d)-(f) Rotation curves predicted by our binary BEC DM model (solid lines)  and the observational data of the three galaxies (triangles) \cite{Harko,Harko1}.}
\label{fig1} 
\end{figure*}

The behavior of the density, the mass, and the tangential velocity of the DM halo as a function of dimensionless radial coordinate, $q r$, for different values of interspecies interaction 
is displayed in Figs.~\ref{fig1} (a),  \ref{fig1} (b) and \ref{fig1} (c). 
We see that $n_{\text{DM}}(r)/n_0$, $M_{\text{DM}}/M_{\odot}$, and $v_{\text{tg}}/v_R$ increase with $g_{12}/g$.

The mass of the atom in the binary mixture can be obtained from the radius of the DM of Eq.~(\ref{DMRad}) as: 
$m= \left(\pi^2 \left(\delta g/g\right)_+ \hbar^2 a / G R_{\text{DM}} \right)^{1/3}$.
For example for $a \approx 10^{6}$ fm, and $g_{12}/g=0.9$, corresponding to the values of the $s$-wave scattering length and interspecies interactions observed in terrestrial Lab experiments,
the mass is of the order of 1.79 ev, while it reduces to 1.49 ev for $g_{12}/g=0.1$. 
These values fulfill the limit of the mass of the condensate particle from cosmological considerations,  $m<1.87$ ev.

A comparison between the velocity of the rotation curves predicted by our binary BEC model and the observational data of 
three galaxies, namely  F 563-V2, DD 01546, and F 583-4 \cite{Harko,Harko1} is made in Figs.~\ref{fig1}(d)-\ref{fig1}(f).
We see that the data and the fit curve are consistent revealing that the binary BEC model is pertinent for describing the DM and its structural properties.

The observation of such a binary DM not only provides diverse and complex structure formation with the diversity of DM halos but also may give a unique opportunity 
in reproducing the observed diversity of dwarf galaxies in analogy with multiple-field fuzzy DM \cite{Berman,Luu1, Pazo, Luu}.

\section{Quantum droplet dark matter} \label{SBDM}

In this section, we check the existence of QD2M and examine its properties at zero temperature.
In the model considered here, the quantum effects are styled on interactions with a dark sector vector field, which
can be achieved in scalar fields with a Coleman-Weinberg effective potential \cite{CW}.
Similarly to most fuzzy DM models, the droplet DM has particularly small masses and couplings \cite{Moss}.

It is well known that if $\left(\delta g/g\right)_+ <0$, the mean-field term provides a term $\propto n^2$ leading to a collapse of a homogeneous state.
The inclusion of the beyond-mean-field LHY term provides an additional repulsive term  $\propto n^{5/2}$ balancing
the attractive mean-field term, allowing quantum fluctuations to stabilize mixture droplets against collapse.
As a consequence, the pressure per particle (\ref{ThPr2}) takes the form: $P = - \left(\delta g/g\right)_+ g n^2 + (128/15)  g \sqrt{a^3/\pi}
  \sum_{\pm}\left(\delta g/g\right)_{\pm}^{5/2}n^{5/2}$.
At zero pressure (or equivalently the energy per particle takes a minimum), the Bose mixture is self-bound and it remains at the equilibrium density which is given by 
$n_{\text{eq}} a^3=\left(25 \pi/16384 \right) \left(\delta g_+/g\right)^2$ \cite{Petrov}. \\
Upon inserting expressions of $\tilde{n}$ and $\tilde{m}$ defined in Eqs.~(\ref{NCD}) and (\ref{MCD}), respectively  up to leading order into Eq.~(\ref{GGPE}), 
we then obtain the following GGPE: 
\begin{align} \label{GGP1}
i\hbar \frac{d\Phi} {dt} &= \bigg[- \frac{\displaystyle\hbar^2}{\displaystyle 2m} \nabla^2+V_g(\mathbf r) + (\delta g)_+ n  \\
&+ \frac{32}{3} \,g  n \sqrt{ \frac{n a^3}{\pi}} \sum_{\pm} \left(\frac{\delta g}{g}\right)_{\pm}^{3/2}\bigg]\Phi. \nonumber
\end{align}
Equation (\ref{GGP1}) is valid when the spatial variations of the condensate density  are small over the extended healing length, $\xi$, validating a description of a locally homogeneous system.
It is important since it describes the distribution of DM particles in the highly dense regions such as the galaxy core. 
For $V_g=0$, Eq.~(\ref{GGP1}) reduces to the standard GGPE governing the static and the dynamics of a self-bound droplet \cite{Petrov,Boudj2}.

Assuming that the out-of-phase motion between the species is not important implies that 
$\Phi(r,t)= \sqrt{n_{\text{eq}}} \phi(r,t)$ \cite{Petrov}, where $\phi (r,t)$ is a scalar wavefunction common to both species.
Then, the dimensionless parameters : $\tilde{N}= N/(n_{\text{eq}} \xi^3)$, $\tilde{r}=r/\xi$, and $\tilde{t}=t/\tau$, where
$\xi= \sqrt{6 \hbar^2/ \left(m |\delta g | n_{\text{eq}}\right)}$, and $\tau = 6 \hbar/\left(|\delta g| n_{\text{eq}}\right)$, result in the following dimensionless  GGPE (\ref{GGP1}):
\begin{align} \label{GGP2}  
i\frac{d\phi} {d\tilde t}= \bigg( -\frac{1}{2} \nabla^2_{\bf \tilde r} +V_g-3|\phi|^2+ \frac{5}{2} |\phi|^3  \bigg)\phi,
\end{align}
which permits a quantitative description of the QD2M using analytical and numerical tools.

\subsection{Hydrodynamic approach} \label{SBDM}

Aiming to derive a differential equation that describes the QD2M,  we employ the hydrodynamic reformulation  of the GGPE (\ref{GGP2}).
Let us now introduce the Madelung representation which requires to write the droplet wavefunction in the form: 
\begin{equation} \label {eq9}
\phi( \tilde {\mathbf  r},t)=\sqrt{n ( \tilde{\mathbf r},t)} \exp [i S (\tilde {\mathbf r},t)], 
\end{equation}
where $S$ encodes the phase which is real and related to the superfluid velocity as $v= \nabla S$. 
After having substituted Eq.~(\ref {eq9}) into the GGPE (\ref{GGP2})  and separating real and imaginary parts, one obtains 
the continuity and the Euler-like equations, respectively
\begin{equation}  \label{Hyd1} 
\frac{\partial n} {\partial t} +{\bf \nabla}\cdot (n \mathbf v)=0, 
\end{equation}  
and 
\begin{align}  \label{Hyd2} 
\frac{\partial {\mathbf v}}{\partial t} &=-{\bf \nabla} \left (V_Q+\frac{1}{2} v^2 +  V_g -3 n +\frac{5}{2} n^{3/2}\right),
\end{align}
where 
$$V_Q=-\frac{1}{2\sqrt{n}} \nabla^2 \sqrt{n}= \frac{1}{4}\bigg [-\frac{(\nabla n)^2}{2 n^2} + \frac{\nabla^2 n }{n}\bigg],$$
represents the so-called quantum pressure which takes into account the Heisenberg uncertainty principle.
Since the fluid is irrotational we can use the property $\nabla (\mathbf v \mathbf v)= 2 (\mathbf v \nabla) \mathbf v$, so Eq.~(\ref{Hyd2}) becomes:
\begin{align}  \label{Hyd3} 
\frac{\partial {\mathbf v}}{\partial t} +(\mathbf v \nabla) \mathbf v+ {\bf \nabla} V_Q+  {\bf \nabla} V_g -3 {\bf \nabla} n + \frac{5}{2} \nabla n^{3/2}=0,
\end{align}
it depends on the strength of the cubic and quadratic nonlinearity terms.

For a static ideal droplet, $\mathbf v=0$, and after applying $\nabla^2$ to both sides of the resulting equation, one obtains
\begin{align}  \label{Hyd4} 
\nabla^2 V_Q+ \nabla^2 V_g -3 \nabla^2 n +V_{\text{LHY}}=0,
\end{align}
where 
\begin{align}  \label{LHYP} 
V_{\text{LHY}}&=\frac{5}{2} \nabla^2 n^{3/2}= \frac{15}{4}\bigg [\frac{1}{2 \sqrt{n}} (\nabla n)^2+ \sqrt{n} \nabla^2 n \bigg]  \\
&= \frac{15}{4} n^{3/2} \bigg[ V_Q-\frac{(\nabla n)^2}{2 n^2}\bigg],  \nonumber
\end{align}
is the LHY potential.\\
In the presence of a gravitational potential which satisfies the Poisson equation $\nabla^2 V_g= 4\pi G n$, 
where $G$ is the gravitational constant (rescaled with above parameters), Eq.~(\ref{Hyd4}) thus reduces to 
\begin{align}  \label{Hyd5} 
V_Q+4\pi G n -3 \nabla^2 n +V_{\text{LHY}}=0.
\end{align}
In the Thomas-Fermi (TF) regime, the quantum potential, $V_Q$, can be neglected and in the absence  of the LHY quantum fluctuations ($V_{\text{LHY}}=0$), Eq.~(\ref{Hyd5}) admits the exact solution
$n({\tilde r}) \sim \sqrt{3} \sinh ({\tilde r}/\sqrt{3}) / {\tilde r}$, which is divergent as ${\tilde r} \rightarrow \infty$. 
This confirms that the LHY correction is crucial for the formation of a stable QD2M.

\subsection{Profiles}

\begin{figure*}
	\includegraphics[scale=0.55]{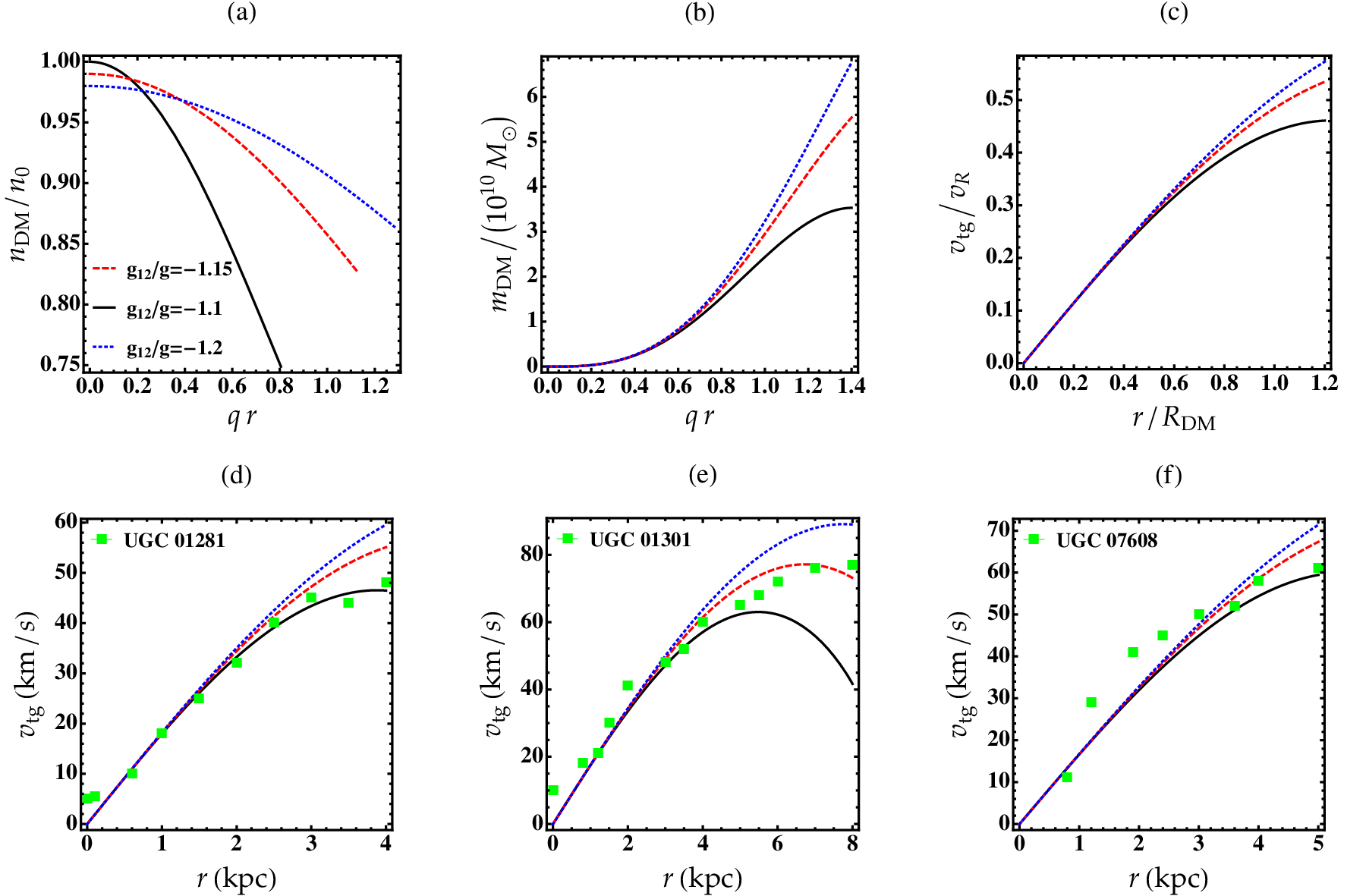}
		\caption{(a) Normalized density profile, $n_{\text{DM}}/n_0$, and (b) Mass distribution of a quantum droplet dark matter for different values of the interspecies interaction.
(c) Normalized tangential velocity, $v_{\text{tg}}/v_R$ as a function of the radial distance $r/R_{\text{DM}}$ of a galaxy consisting 
of a quantum droplet matter halo for different values of the interspecies interaction. 
(d)-(f) Rotation curves predicted by our QD2M model (solid lines)  and the observational data of the three galaxies (squares) \cite{Harko,Harko1}.}
\label{fig3} 
\end{figure*}

In the presence of $V_{\text{LHY}}$, Eq.~(\ref{Hyd5}) cannot be integrated due to the $(2/{\tilde r})\nabla n$ term. 
However, to solve this equation we should search for a spatially localized solution by requiring that
$n \propto e^{- \alpha \tilde r}/{\tilde r}$ as $\tilde r \rightarrow \infty$, and  a smooth solution at ${\tilde r}=0$ by imposing the condition $\nabla n=0$ at ${\tilde r}=0$ (see e.g. \cite{Must}). 
The numerically obtained profiles that satisfy our requirements  are displayed in Fig.~\ref{fig3}.

The results highlight that the interspecies interaction enhance the density QD2M results in a larger radius as seen Fig.~\ref{fig3} (a)
indicating that the QD2M would behave like very large CDM particles.
The central density is decreasing monotonically with the radial distance similarly to the BEC-DM matter model.
From Fig.~\ref{fig3} (b), it is easy to see that the distribution of quantum droplet  DM mass is slowly increasing  with the increasing distance from the center 
notably for relatively large interspecies interaction ($\delta g/g=-1.1$) in contrast with binary BECs DM studied above.
However, it increases rapidly for small $\delta g/g$ confirming the existence of the DM which may support larger maximum stellar masses.
As shown in Fig.~\ref{fig3} (c), the velocity of the rotation curves obtained with our QD2M model preserves the same behavior as in the binary BEC DM results predicted above. 

In order to check the adequacy of our theoretical predictions with observational tangential velocities, we perform  three bulgeless galaxy rotation curve fits
and assume a constant stellar mass-to-light ratio $r_{\text {disk}}=0.5$ (see e.g. \cite{Delgad,Lell}).
Therefore, the free parameters of the rotating QD2M model are the radius and central density. 
The admissible values are obtained through $\chi^2= \sum_i \left(v_{\text {obs},i}- v_{\text{DM},i} \right)^2/v_{\text{DM},i}$, where 
$v_{\text {obs},i}$ are the values of observed velocity \cite{Harko, Freund}, which must fulfill the condition $\chi^2<1$, providing the desired rotation curves.
The resulting best fit rotation curves of three galaxies (UGC01281,  UGC01310, and UGC07608) \cite{Harko,Harko1} are shown in Figs.~\ref{fig3} (d)-(f).
The fitted values of the parameters together with the values of $\chi^2$ for the above sample galaxy group are summarized in Table \ref{table:1} to estimate the goodness-of-fit.   
We find good overall agreement between our theoretical predictions and the observational data for all probed galaxies as can be seen
from Figs.~\ref{fig3}(d)-\ref{fig3}(f)  and Table \ref{table:1} ($\chi^2<1$).

\begin{table}[h!]
\begin{center}       
\begin{tabular} { cccc cc} 
 \hline \hline\\
                       & Galaxy  \;\;          & $R (\text{kpc})$ \; \;         & $n_0 (10^{-24} \text{g/cm}^{3}) $ \;\;\;  & $ \chi^2$ \\ 
 \hline\\
                       & UGC01281      & 4.05                      & 1.6                                           & 0.2\\  
                       & UGC01310      & 7.9                        & 1.03                                               &0.8 \\
                       & UGC07608      & 5.1                        & 3.8                                               & 0.3 \\
 \hline\hline
\end{tabular}
\end{center}
The best-fit values of the radii and central densities for the three  galaxies shown in Fig.~\ref{fig3}.
\label{table:1}
\end{table}

\subsection{Stability}

Let us now study the linear dynamical stability of the observed QD2M with velocity ${\bf v}=0$, described by the above quantum hydrodynamic equations. 
This is a generalization of the classical Jeans problem \cite{Jean,Khlo, PH1,Harko3} to a static self-bound quantum fluid.
Note that the Jeans instability of an infinite homogeneous self-gravitating BEC governed by the usual GPE has been analyzed in a static and in an expanding universe 
(see e.g. \cite{Khlo, PH1,Harko3}).

We assume small fluctuations around a homogeneous background in an equilibrium state, $n(\tilde {\mathbf r},t)= n(\tilde {\mathbf r})+ \delta n(\tilde {\mathbf r},t)$, 
where  $\delta n(\tilde {\mathbf r},t)/n (\tilde {\mathbf r})\ll1$. 
Shifting the phase by $-\mu t$, where $\mu$ is the chemical potential of the droplet, and linearizing the hydrodynamic equations (\ref{Hyd1})-(\ref{Hyd2})
with respect to $\delta n$ and $\nabla S$, we then obtain
\begin{equation}  \label{chimB}
\mu=-\frac{1}{2} \frac{\nabla^2 \sqrt{n}} {\sqrt{n}}+ V_g -3n+ \frac{5}{2} n^{3/2}.
\end{equation}
The first-order terms provide equations for the density and the velocity as
\begin{equation}  \label{hydo3}
\frac{\partial \delta n}{\partial t} +{\bf \nabla}.(n \mathbf v)=0,
\end{equation}
and 
\begin{equation}  \label{hydo4}
\frac{\partial \mathbf v}{\partial t} =-{\bf \nabla} \delta\mu,
\end{equation}
Taking the time derivative of  Eq.~(\ref{hydo3}) and inserting  Eq.~(\ref{hydo4}) into the resulting equation, we get
\begin{equation}  \label{breth}
 \frac{\partial^2 \delta {n}} {\partial t^2} ={\bf \nabla}. (n {\bf \nabla} \delta\mu),
\end{equation}
where $\delta\mu$ can be derived from Eq.~(\ref{chimB}).
Expanding the solutions of Eq.~(\ref{breth}) into plane waves $\delta n \propto e^{i ({\mathbf k. \tilde {\mathbf r}}- \omega t)}$ and employing $\nabla^2 \delta V_g= 4\pi G \delta n $,
we find for excitation spectrum:
\begin{equation}  \label{BogDR}
\omega= \sqrt{ \frac {k^4}{4}+ n \left(-3+\frac{15}{4} \sqrt{n} \right) k^2+4\pi Gn}.
\end{equation}
In the nongravitational limit, $G=0$, Eq.~(\ref{BogDR}) reduces to 
$$\omega= k\sqrt{ k^2/4+ n \left(-3+15\sqrt{n}/4 \right)}.$$
For small momenta, $k \rightarrow 0$, the excitations are sound waves  $\omega= c_s k$, where $c_s =\sqrt{ n \left(-3+15\sqrt{n}/4\right)}$ is a dimensionless sound velocity.
Importantly, a phonon instability appears when the density reaches a certain critical value, $n_{\text{cr}}=4/5$ which is followed by a transition to an inhomogeneous gas
for $ n <n_{\text{cr}}$, where the droplet loses its flat-top shape.
In the absence of the LHY term, one recovers the standard Bogoliubov dispersion relation for BEC, $\omega= k\sqrt{ k^2/4+ n}$.

In the classical limit (TF regime), the first term in Eq.~(\ref{BogDR}) can be neglected, thus the dispersion relation simplifies to
\begin{equation}  \label{BogCl}
\omega= \sqrt{ c_s^2 k^2+4\pi Gn}.
\end{equation}
Remarkably, for $c_s^2<0$, the QD2M is stable for these modes which is indeed natural since the droplet requires a negative chemical potential.
This can be attributed to the balance between gravity and quantum LHY corrections. 
However, for $c_s^2>0$, the system is unstable.  This is in contrast with the well-known Jeans instability for the usual BEC DM (see e.g. \cite{Jean,Khlo, PH1,Harko3}),  
which requires a negative gravity and a positive chemical potential (or equivalently sound velocity).
If $c_s=0$, the dispersion relation reads $\omega= \sqrt{4\pi Gn}$  and the oscillation evolves with time as $\delta n \propto e^{-i \sqrt{4\pi Gn} t}$.

\begin{figure}
	\includegraphics[scale=0.8]{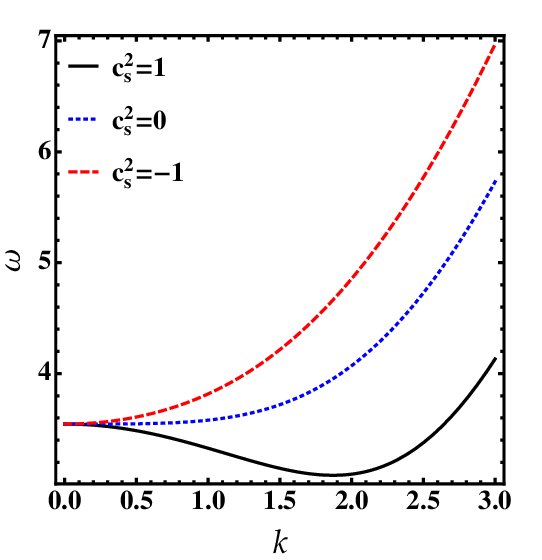}
		\caption{Dimensionless dispersion relation for QD in the general case for different values $c_s$.}
\label{fig2} 
\end{figure}

In the general case, the dispersion relation (\ref{BogDR}) can be rewritten as:
\begin{equation}  \label{BogJeans}
\omega= \sqrt{ (k^2-k_J^2) \left[c_s^2+ \frac{1}{4} \left (k^2+k_J^2\right)\right]},
\end{equation}
where $k_J^2= 2c_s^2 \left[-1+ \sqrt{1- 4\pi Gn/c_s^4} \right]$ is the Jeans-like wavenumber, corresponding to $\omega=0$.
It follows from Eq.~(\ref{BogJeans}), that for $k=k_J$, the Bogoliubov frequency vanishes and hence the perturbation is stationary. 
For $c_s^2 <0$, one has $\omega^2>0$, then the QD under the gravitational potential is always stable. 
If $c_s^2 >0$ the system is stable only for $k>k_J$.
The behavior of $\omega$ is shown in Fig.~\ref{fig2}.
Evidently, the interplay between the  gravitational potential, the repulsive LHY quantum corrections and
the attractive mean-field interactions plays a crucial role in the behavior of the dispersion relation and thus in the stability of the system.

\section{Time evolution of QD2M}

In this section, we analyze the dynamical properties of QD2M parameters using variational super-Gaussian ansatz which incorporates both
the time evolution of the DM radius and its breathing oscillations.

The total energy associated with the hydrodynamic system (\ref{Hyd1})-(\ref{Hyd2}), or equivalently with the GGPE (\ref{GGP2}), reads
\begin{align}\label{engyfun}
E=\int_{0}^\infty \bigg [\frac{1}{2} \left|\nabla \phi \right|^2+\frac{1}{2} V_g |\phi|^2  -\frac{1}{2} |\phi|^4+ \frac{2}{5} |\phi|^5\bigg] d^3{\tilde r},
\end{align}
which gives insights on the stability of the droplet.\\
The Lagrangian density generating Eq.~(\ref{GGP2}) reads
\begin{align}\label{Lagr}
{\cal  L}= 	\frac{i}{2} \left( \phi  \dot{\phi}^*- \phi^*\dot{\phi}  \right)+\frac{1}{2} \left|\nabla \phi \right|^2+\frac{1}{2} V_g |\phi|^2  -\frac{1}{2} |\phi|^4+ \frac{2}{5} |\phi|^5,
\end{align}
where $\dot{\phi}= d \phi/dt$.
In order to calculate analytically the density and mass profiles of QD2M, we use a super-Gaussian trial function (see e.g. \cite{Asma,Otajo,Indj} and references therein)
\begin{eqnarray}\label{SGauss}
\phi(r)= A \exp\bigg[-\frac{1}{2}\left(\frac{\tilde r}{R}\right)^{2 \gamma}+i \beta {\tilde r}^2\bigg],
\end{eqnarray}
where $\gamma$ is the super-Gaussian index which can be determined from the parameters of a stationary solution, 
$A$ is the amplitude, $R$ is the droplet width, and $\beta$ accounts for the phase of the  droplet.
The normalization condition, ${\tilde N}=4\pi \int_{0}^{+\infty} {\tilde r}^2 d{\tilde r} |\phi|^2$,  yields
 \begin{align}  \label{Norsup}
A= \frac{1} {2 R^{3/2} } \sqrt{ \frac{3 \tilde {M}/\pi} {\Gamma \left(1+3s\right)}},
 \end{align}
where  $\Gamma[z]$ is the gamma function, and $s=1/(2\gamma)$.
Here we used the fact that the total mass of the DM halo, $\tilde {M}$, is  of the order of the number of the DM particles ${\tilde N}$ (unit of mass).

\begin{figure}
	\includegraphics[scale=0.45]{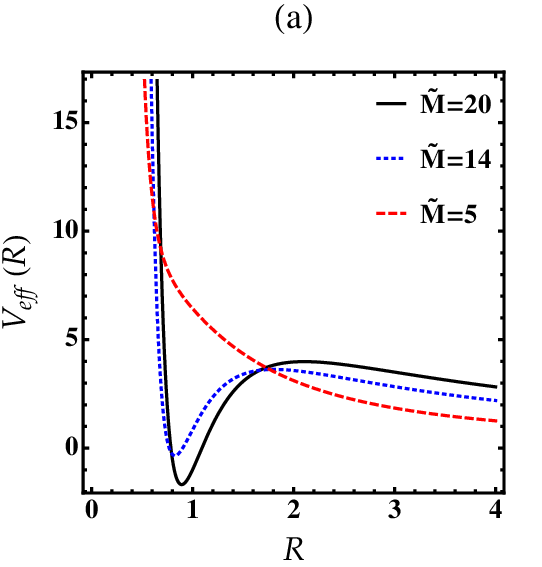}
	\includegraphics[scale=0.45]{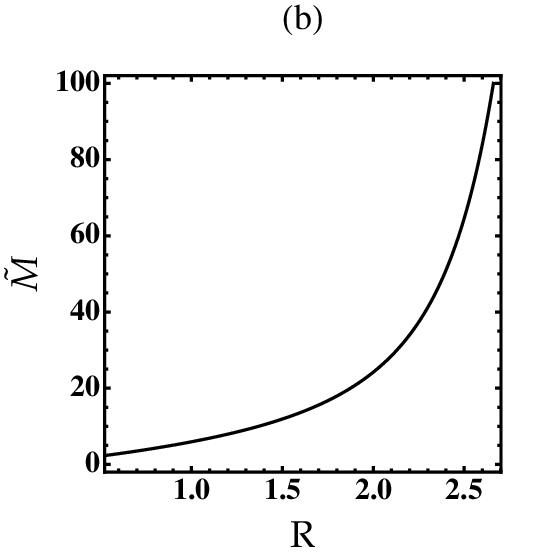}
	\includegraphics[scale=0.45]{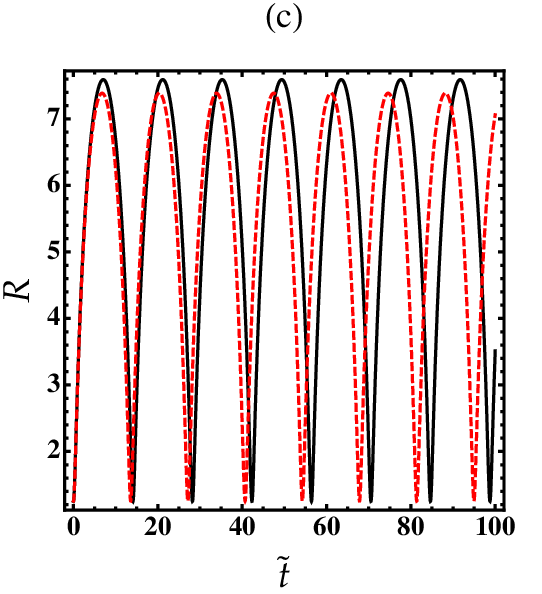}
	\includegraphics[scale=0.45]{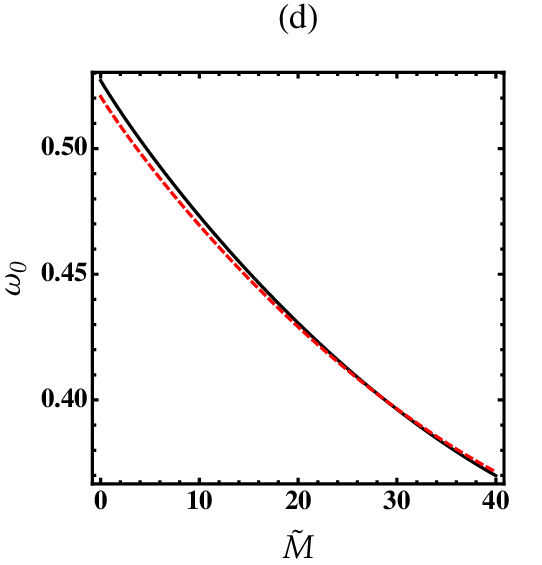}
		\caption{ (a) Effective potential of QD2M for different values of $\tilde {M}$.
(b) Variation of the dimensionless mass, $\tilde {M} (R)$, of QD2M as a function of the radius, $R$. (c) Time evolution of the radius (oscillations) of QD2M. 
(d) Frequency of the breathing modes, $\omega_0$, as a function of the mass $\tilde {M}$. 
Black solid lines correspond to variational calculation. Red-dashed lines correspond to numerical simulations.
Numerically, the frequency can be evaluated via $\omega_0=2\pi/T$, where  $T$ is the average period of  the amplitude oscillation.
Parameters are: $G=1$, and $\beta = 0.01$.}
\label{Var} 
\end{figure}

The averaged Lagrangian, $L= 4\pi \int_0^{\infty} {\cal L} {\tilde r}^2 d{\tilde r}$, can be computed by introducing the trial function (\ref{SGauss}) 
into Eq.~(\ref{Lagr}), and using Eq.~(\ref{Norsup}):
\begin{align}\label{Lagr}
L= 	\tilde {M} \dot \theta+ \tilde {M}\frac{3 \Gamma \left(1+5s\right)}{5\Gamma \left(1+3s\right)} R^2(2\beta^2+\dot \beta)+f(R,\tilde {M},s),
\end{align}
where
\begin{align}
		f(R,\tilde {M},s)=&\frac{3 \sqrt{3}\, 2^{3s-1}  \tilde {M}^{5/2}}{\pi ^{3/2}  5^{1+3s}\Gamma \left(1+3s\right)^{3/2} R^{9/2}}
+\frac{3 \tilde {M} \Gamma \left(2+s\right)}{4s \Gamma \left(1+s\right)R^2} \nonumber \\
&+\frac{3 \tilde {M}^2 G}{\ 2^{3(1+s)}\Gamma \left(1+3s\right) R}-\frac{3  \tilde {M}^2}{\pi   2^{2+3s}\Gamma \left(1+3s\right) R^3}. \nonumber
	\end{align}
The dynamics of the radius of QD2M can be obtained from the Euler-Lagrange equation:
	\begin{equation}\label{Va4}
	\tilde {M} \ddot{R}=-\frac{d V_{\text{eff}}(R)}{d R},
	\end{equation}
where the effective potential for the oscillations of the droplet's width is given by:
\begin{align}\label{Veff}
		V_{\text{eff}}(R)=\frac{5 \Gamma \left(1+3s\right)}{3\Gamma \left(1+5s\right)} \frac{f(R)}{\tilde M} .
	\end{align}
For fixed parameters, $R$ and $\tilde {M}$, the super-Gaussian index, $\gamma$, can be found from the variational equation $d L/d\gamma=0$ \cite{Asma,Otajo}.
This condition allows us also to determine the critical mass beyond which the QD2M destabilizes.
Equation (\ref{Va4}) indicates that there is a fictive particle with mass $\tilde {M}$ and position $R$ moving in a potential $V_{\text{eff}}(R)$.
The local minimum of this latter corresponds to the equilibrium radius $R_{\text{eq}}$. 
The potential $V_{\text{eff}}(R)$ is plotted in Fig.~\ref{Var} (a). 
As can be seen from the figure, a stable QD2M occurs for larger $\tilde {M}$ where $V_{\text{eff}}(R)$ exhibits a local minimum at $R_{\text{eq}} \simeq 0.9$.
In such a case the amplitude of the droplet reaches the TF limit. Reducing the $\tilde {M}$, the droplet becomes unstable and evaporates eventually.
The spatial variation of the dimensionless mass, $\tilde {M} (R)$ corresponding to the equilibrium state is represented in Fig.~\ref{Var} (b). 
It is characterized by a monotonically increasing function of $R$. At distances $(R>R_{\text{eq}})$, it increases sharply with $R$.
The dynamics of the QD2M radius is captured in Fig.~\ref{Var} (c). We see that it oscillates periodically around the equilibrium state during the time evolution.

The radius variation around the equilibrium state signals the influence attributed to the breathing mode oscillation, and
its frequency can be computed by linearizing Eq.~(\ref{Va4})  around the equilibrium solutions, $R (t)=R+\delta R(t) $ 
(with $\delta R\ll R$) and expanding the effective potential (\ref{Veff}) into a Taylor series.  Keeping only leading terms in $\delta R(t)$, we obtain
\begin{equation}\label{Bmodes}
		\omega_0= \sqrt{\frac{1}{\tilde M}\frac{ \partial^2 V_{\text{eff}}(R)}{\partial R^2} \bigg|_{R=R_{\text{eq}}}}.
\end{equation}
A steady state is linearly stable only for $\omega_0>0$.
As shown in Fig.~\ref{Var} (d),  the frequency of the breathing modes is decreasing with the mass $\tilde {M}$. 
For QD2M with a halo mass $M_{\text{DM}}/M_{\odot} = 10^{11}$, $R_{\text{DM}}=10$ kpc $=3.08\times 10^{22}$cm, and
$a \approx 4.71\times 10^{-7}$ cm, we find for the period of the oscillations $T=2\pi/\omega_0 \approx 5.7$ Gyrs.

It is apparent from Fig.~\ref{Var} that the variational method excellently agrees with simulations and captures the dynamics of the QD2M.

\section{Conclusion and outlook} \label{concl}

In this paper we exhaustively investigated the existence of QD2M of Bose mixtures, highlighting the role of the LHY quantum corrections 
in the formation and stability of this object.
Within the framework of the HFB theory we derived an extended EoS which becomes polytropic with index 1 for a symmetric binary BEC.
In this case, we calculated analytically the density and mass profiles and the radius of DM for different values of the interspecies interactions 
which include appropriate corrections to the results obtained from a single BEC DM model.
In order to test our binary BEC DM model we compared the tangential velocity equation with the data of dwarf galaxies, 
and good agreement is found, offering a rich structure formation of DM haloes.

By introducing the Madelung representation of the droplet wavefunction, it follows that the dynamics of the DM can be governed by a set of coupled hydrodynamical equations
showing the appearance a barotropic like-flow.  
The numerical solutions of the resulting EoS reveal that the density and mass profiles and the radius of the QD2M are sensitive to the interspecies scattering length, and the LHY strength.
In addition, we studied  the stability properties of the QD2M halo in the homogeneous case (flat-top regime) by linearizing the hydrodynamic equations
leading to perturbation equations that describe the collective excitations of the droplet. 
By means of a variational super-Gaussian trial function, the equilibrium properties, the evolutionary equation, and the frequency of the breathing oscillation of QD2M
have been adequately captured.

Our model not only circumvents the known cusp-core problem, and enables us to obtain all the relevant quantities in terms of 
the observable parameters such as the dark halo mass, the radius of the galaxy, and the tangential velocity in a straightforward manner,
but also shows that QDs could be good DM candidates excellently fit  to the observational data.

\section*{Acknowledgments}

We thank Tiberiu Harko, Pierre-Henri Chavanis and Ian Moss for valuable discussions and useful comments about the paper.

\end{document}